\documentclass[11pt]{revtex4}
\usepackage[cp1251]{inputenc}
\usepackage[english,russian]{babel}
\usepackage{indentfirst}
\usepackage{graphicx}
\usepackage{amsmath}
\usepackage{amssymb}
\usepackage{bm}
\input{epsf}

\newcommand{\vv}{\mathbf}

\begin{document}

\noindent UDK \,524.57

\noindent PASC \,98.38.Cp, 98.38.Am, 95.30.Qd

\title{ON SATURATION OF BETATRON ACCELERATION OF DUST PARTICLES BEHIND SHOCK 
FRONTS}

\author{\firstname{L.~V.}~\surname{Kostyukova}}
\affiliation{Faculty of Physics of Southern Federal University, 
Rostov on Don, Russia}

\author{\firstname{V.~V.}~\surname{Prudskikh}}
\affiliation{Faculty of Physics of Southern Federal University, 
Rostov on Don, Russia}

\author{\firstname{Yu.~A.}~\surname{Shchekinov}}
\affiliation{Faculty of Physics of Southern Federal University, 
Rostov on Don, Russia} \affiliation{Special Astrophysical Observatory of RAS,
 \protect\\
 Nizhny Arkhyz, Karachai-Cherkessiya, Russia}

\received{22.06.2009 Ј.}
\revised{10.08.2009 Ј.}
\begin{abstract}
Possible saturation of betatron acceleration of dust particles behind 
strong shock fronts from supernovae is considered. It is argued that the 
efficiency of the nonthermal dust destruction should be substantially lower 
than the value estimated from a traditional description of betatron 
acceleration of dust grains behind radiative shock waves. The inhibition 
of the nonthermal destruction can be connected with the mirror instability 
developed in the dust component behind strong shocks with the velocity 3 times 
exceeding the Alfv\'en speed. The instability develops on characteristic time scales 
much shorter the age of a supernova remnant, thus its influence on the 
efficiency of dust destruction can be substantial: in the range of shock 
velocities 100~km s$^{-1}<v_s<300$~km~s$^{-1}$ the destruction efficiency 
can be an order of magnitude lower that normally estimated. 
\end{abstract}

\maketitle

\section{INTRODUCTION}

At present it is commonly accepted that dust grains are efficiently destroyed 
behind shock fronts from supernovae~[1, 2]. A rough estimate of the dust destruction 
rate in the Galaxy can be obtained from the following simple consideration~[2]: 
the total mass of the dust destroyed in unite time is 
$\dot M_d= \nu_{SN}\eta M/\zeta$, where $\nu_{SN}$ is the Galactic supernova 
rate, $\eta$ is the efficiency of dust destruction by a shock with a given 
velocity, $M$ is the mass of supernova shell with a given velocity; the kinetic 
energy of the shell is assumed by order of magnitude to be equal to the 
explosion energy $E\sim Mv^2/2$. It is commonly assumed that dust grains most 
efficiently are destroyed by the shock waves with the velocities $v\geq
100$~km~s$^{-1}$ --- at lower velocities the destruction efficacy drops 
exponentially ~[1]. The fraction of destroyed dust behind the shock with the 
velocity $v_{s,0.4}=100$~km~s$^{-1}$ is around $\eta\simeq 0.4$, hence the 
total rate of the dust mass decrease in the Galaxy is estimated as 

\begin{equation}
\dot M_d\sim 0.8\nu_{{SN}}\frac{E_{{SN}}}{\zeta v_{s,0.4}^2}
\end{equation}
where $E_{SN}=10^{51}$~erg is the explosion energy, 
$\zeta=100{-} 200$ is the mean gas-to-dust mass ratio. The characteristic time 
of dust destruction is 

\begin{equation}
\tau_{d,d}=\frac{M_d}{\dot M_d}=1.25\frac{M_gv_{s,0.4}^2}{E_{SN}\nu_{SN}} \sim 10^8\textrm{~year},
\end{equation}
where $M_g=5\times 10^9~M_{\odot}$, the gas mass in the Galaxy, 
the suparnova rate is assumed to be $\nu_{SN}=1/100$~year$^{-1}$.

It is obvious that this value of $\tau_{d,d}$ is unacceptably short in comparison 
with the time of dust formation (replenishment) in the Galaxy. Indeed, 
the total rate of dust injection into the interstellar medium (ISM) from 
stellar sources (including supernovae) is ${\leq} 3\times
10^{-11}~M_{\odot}$~pc$^{-2}$~year$^{-1}$~[3-5]. Assuming now the total dust mass 
in the Galaxy $\sim 5\times 10^7~M_{\odot}$, one can estimate the replenishment 
rate is $\tau_{d,f}\sim 10^{9}$~years. In such conditions the total dust mass 
in the Galaxy must be an order of magnitude lower than the observed value, 
i.e. $\sim 5\times 10^5~M_{\odot}$.

Moreover, it is well known that the abundances of refractory elements in gas 
phase of the ISM are always substantially lower than solar values. In other 
words such elements are ``depleted'' in gas phase. This depletion is connected 
with their condensation on the surface of dust grains. Even in the diffuse 
rarefied phase ($n\simeq 0.05$~cm$^{-3}$) the depletion factor can be as high as 
90$\%$ for Fe, Ca, Ti, i.e. the fraction of these elements in gas phase is 
10$\%$ or lower, while in more dense regions ($n\simeq 3$~cm$^{-3}$) the fraction 
of Fe in gas phase is around 1$\%$, and the fraction of Ca and Ti is onlu 
0.1$\%$~[6,7]. This means that destruction of dust (and as a consequence 
removal of metals from condensed phase into gaseous) should be considerably 
slower process than their formation, i.e. condensation of metals on to the 
solid (dust) phase. In particular, for the diffuse ISM the following 
interrelation between the characteristic dust destruction $\tau_{d,d}$ and 
formation $\tau_{d,f}$ times: $\tau_{d,f}=0.1\tau_{d,d}$. In general, 
the condition $\tau_{d,f}\ll \tau_{d,d}$ has to be always fulfilled, though 
at present the mechanisms able to maintain this condition are unknown. 

One of the possibilities is connected with that a considerable dust mass 
is confined in dense atomic and molecular clouds. When a shock wave falls 
on to a dense cloud its velocity inside the cloud decreases as 
$v_{s,c}\sim v_{s,i}\sqrt{\rho_i/\rho_c}$, where indices $c$ and $i$ belong 
to the cloud and the intercloud medium, respectively. Even for HI clouds 
with density only one or one and a half orders higher than in the intercloud 
medium the shock velocity decreases by factor ~3--5~, so that dust remains 
undestroyed behind such shock waves. Therefore, only dust in the diffuse 
intercloud gas is destroyed by supernovae shocks. With this circumstance 
accounted the above estimate can be rewritten as follows 

\begin{equation}
 \dot M_d\sim 0.8\mu_{i}\nu_{SN}\frac{E_{SN}}{\zeta v_{s,0.4}^2}.
\end{equation}
Here $\mu_{i}=\rho_i(1-f_c)/\rho_cf_c$ is the mass fraction contained in the 
intercloud medium, $f_c$ is the volume filling factor of clouds: 

\begin{equation}
f_c=\frac{4}{3}\pi V_{\textrm{MC}}^{-1}\int R_c^3\frac{dN}{dM}dM,
\end{equation}
where the cloud mass spectrum $dN/dM\propto M^{-1.7}$ in the mass range 
$10^{-4}~M_{\odot}<M<10^5~M_{\odot}$~is assumed [8, 9]; the estimate gives 
$\mu_i\sim 0.1$. As a result, the dust destruction time increases as 
$\tau_{d,d}\sim 2\times 10^8/\mu_i$ and becomes comparable to the Galactic dust 
production time. At the same time the necessary interrelation 
$\tau_{d,f}\ll \tau_{d,d}$ remains unexplained.

Apparently, one has to assume that in such conditions condensation of metals on to 
solid (dust) phase should occur directly in the ISM. A low abundance of 
deuterium in gas phase of the local interstellar medium seems to favor this 
conclusion. It is argued in ~[10] that the observed relative abundances of 
deuterium and oxygen indicate that deuterium is frozen-out on dust grains: 
the characteristic time for deuterium to freeze out is estimated in ~[10] 
to be around several Myr in the conditions of diffuse interstellar clouds. 

The latter means apparently that dust grains {\textit{escape destruction}} 
by supernovae shocks. Indeed, all known data about the depletion of 
deuterium belongs to the nearest vicinity of Sun restricted by distances 
of 100--150~pc. The characteristic interval between successful explosions of 
supernovae in this volume is around $5\times 10^5$~year~[11] -- an order 
of magnitude shorter than the time of deuterium freezing on the dust surface. 
Therefore deuterium can be depleted in gas phase of the local ISM provided 
only that SNe shocks {\textit{do not destroy}} dust grains. 

In this paper we present arguments favoring that plasma instabilities 
behind shock waves prevent an efficient acceleration of charged dust particles 
by betatron mechanism, and thus inhibit their destruction. An essential 
circumstance here is that the dust component can change its dynamics only on 
times corresponding to the inverse dust cyclotron frequency $\omega_d$. 
This requires the presence of plasma oscillations with frequences close to 
$\omega_d$ in the whole spectrum, which seems plausible only under description of 
the dust as a separate dynamical component. Therefore, in this paper we describe 
instabilities in the interstellar plasma behind strong shock waves caused 
by the presence of the charged dust component. 

In Sect. 2 we formulate the equation of motions of a three-fluid (ions, 
electrons and charged dust) interstellar plasma, in Sect. 3 we derive the 
dispersion equation for small oscillations in such a plasma, in Sect. 4 
the dispersion equation is analyzed and the conditions for instabilities are 
discussed, Sect. 5 summarizes the results. 
 

\section{EQUATIONS OF MOTION} 

An efficient destruction of dust particles behind shock fronts is determined 
by their betatron acceleration in the course of plasma compression on radiative 
stages~[2, 12]. Namely, during a radiative cooling and an accompanying 
compression of plasma magnetic field behind the shock front grows as 
$B\propto \rho$, what in turn leads to a proportional increase of the 
square of the transversal momentum of dust particles $p^2_{\perp}\propto B$. 
Consequently the efficiency of dust destruction increases by more than an order 
of magnitude for the shocks with velocities 100~km~s$^{-1}
<\linebreak <v_s<300$~km~s$^{-1}$. In these estimates the structure of the 
post-shock flow, in particular, the magnetic field being parallel to the 
front, is assumed implicitly to be stable and invariable over the whole 
post-shock region. At the same time, one can think that growing plasma 
instabilities are able to violate the symmetry of the flow and moreover to 
make the flow irregular. As a result, plasma regions with a decreased magnetic 
field and the correspondingly decreased dust velocities can arise, which result 
in an inhibition of the dust destruction efficiency. 

One of the instabilities which can cause violation of a regular magneto-hydrodynamic 
flow is the mirror instability of an anisotropic plasma with the transversal 
temperature exceeding the parallel one: $T_{\perp}\gg
T_{||}$. In the conditions behind shock fronts anisotropy of the dust component 
emerges naturally: after crossing a shock front with a parallel magnetic field 
a dust grain gains the transerval velocity in the restframe connected with the front 
close to the shock velocity, while the longitudinal velocity remains equal to 
a much lower velocity of random dust motions in the ISM. Below we consider 
a possibility for the mirror instability to develop in such conditions. 

Let us describe the interstellar plasma by the equations~[13] 

\begin{equation}
\partial_tn_\alpha+\nabla\cdot (n_\alpha {\mathbf{u}}_\alpha)=0,
\end{equation}
where $\alpha=i,e,d$ corresponds to the ions, electrons and dust particles. 
Conservation of the momentum is described by the equations 

\begin{eqnarray}
\partial_t(n_\alpha{\mathbf{u}}_\alpha)&+&\nabla\cdot
(n_\alpha{\mathbf{u}}_\alpha
{\mathbf{u}}_\alpha)= \nonumber\\
={-}\frac{1}{m_\alpha}\nabla p_\alpha &-&\frac{n_\alpha
\tilde{Z}_\alpha e}{m_\alpha} \left(\nabla\Phi-\frac{1}{c}[{\mathbf{v}}_\alpha{\mathbf{B}}]\right),
\end{eqnarray}
which should be complemented by the equations 
\begin{equation}
\Delta\Phi={-}4\pi e\sum \tilde{Z}_\alpha n_\alpha
\end{equation}
and 
\begin{equation}
\nabla\times {\mathbf{B}}=\frac{4\pi e}{c}\sum \tilde
Z_\alpha n_\alpha {\mathbf{v}}_\alpha,
\end{equation}
as well as the equation for $\mathbf{{\rm curl}E}$; here $\tilde
Z_e={-}1$, $\tilde Z_i={+}1$ and  $\tilde Z_d={-}Z_d$, so that in typical 
conditions with negative dust grains the quasi-neutrality equation has the 
form $n_i=n_e+\tilde Z_d n_d$.

Let us consider plasma behind the shock in the comoving (moving with the plasma) 
restframe. Let us consider also the configuration with magnetic field poins 
parallel to $z$-axis (along the shock front), while the wavevector 
has the components $\mathbf{k}=(k_{\perp},0,k_{||})$. In this work we neglect 
the effects connected with perturbations of the front. Therefore our results 
formally are valid far from the front -- stability of the shock front in such 
a multi-component plasma deserves a separate consideration. 

Let us linearize the system~(5)--(8) in projections for each plasma components. 
We will restrict consideration by low-frequency electromagnetic waves with 
the phase velocity satisfying the inequality 

\begin{equation}
v_{Ti}\ll\omega/k\ll v_{Te},
\end{equation}
where $v_{Te}$ and $v_{Ti}$ are thermal velocities of the electrons and ions, 
respectively. For linearized equations equations of motion of the electrons 
one obtaines 

\begin{equation}
\frac{\partial v_x}{\partial
t}={-}\frac{e}{m}E_x-\omega_ev_y-i\frac{k_{\perp}
T_e}{m}\frac{\widetilde{n_e}}{n_{e0}} ,
\end{equation}
\begin{equation}
\frac{\partial v_y}{\partial t}={-}\frac{e}{m}E_y+\omega_ev_x ,
\end{equation}
\begin{equation}
0={-}\frac{e}{m}E_z-i\frac{k_{||}T_e}{m}\frac{\widetilde{n_e}}{n_{e0}}
,
\end{equation}
where $\omega_e=eB_0/mc$ is the electron cyclotron frequency,
$T_e$ is the electron temperature. In equation~(12) the electrons are 
assumed to be inertialess along the field $B_0$,
which corresponds to the right hand side of the inequality~(9). Then 
the continuity equation for the electrons turns out to be equivalent to 
the equation for a perturbation of the electron density derived from the 
latter equation: 

\begin{equation}
\frac{\widetilde{n_e}}{n_{e0}}=i\frac{e}{k_{||}T_e}E_z.
\end{equation}

The ion equations of motion are written as 

\begin{equation}
\frac{\partial V_x}{\partial t}=\frac{e}{M}E_x+\omega_iV_y,
\end{equation}
\begin{equation}
\frac{\partial V_y}{\partial t}=\frac{e}{M}E_y-\omega_iV_x,
\end{equation}
\begin{equation}
\frac{\partial V_z}{\partial t}=\frac{e}{M}E_z,
\end{equation}
where terms with thermal pressure are neglected in virtue of the left 
inequality (9). The continuity equation for the ions couples their density 
perturbations with variations of both longitudinal and transversal velocities: 

\begin{equation}
\frac{\partial}{\partial
t}\left(\frac{\widetilde{n_i}}{n_{i0}}\right)+\frac{\partial
V_x}{\partial x}+\frac{\partial V_z}{\partial z}=0.
\end{equation}

Contrary to the electron and ion components dust relaxes to thermal equailibrium 
much slower because of low frequency of collisions. Therefore one can assume 
that in the region where the elctrons and ions are isotropic the dust component 
is anisotropic. In such conditions in the equations of motion of the dust the 
anisotropic pressure tensor has to be explicitly used. The linearized gradient of 
the pressure tensor is ~[14] 

\begin{eqnarray}
\nabla P=\widehat{x}\left[\frac{\partial
p^{\prime}_{\perp}}{\partial x}+\frac{\partial}{\partial
z}(p_{||}-p_{\perp})\frac{B_x}{B_0}\right]+ \nonumber\\
+\widehat{y}\frac{\partial}{\partial
z}(p_{||}-p_{\perp})\frac{B_y}{B_0}+\widehat{z}\left[\frac{\partial}{\partial
x}(p_{||}-p_{\perp})\frac{B_x}{B_0}+\frac{\partial
p^{\prime}_{||}}{\partial z}\right].
\end{eqnarray}
Here $p_{||}=n_{do}T_{||}$, $p_{\perp}=n_{do}T_{\perp}$, $n_{d0}$
is dust density, $T_{||}$ and $T_{\perp}$ its temperature along and 
perpendicular to the magnetic field, respectively, $p^{\prime}_{||}$, 
$p^{\prime}_{\perp}$ are small deviations of pressure components from 
the equilibrium values. It has to be mentioned that since the field line is 
determined by superposition of the external and the wave field, the pressure 
$p_{||}$ and $p_{\perp}$ in different points should have different projections 
on to $x$ and $z$ axes. Then the motion equations of dust get the form 

\begin{eqnarray}
\frac{\partial w_x}{\partial
t}=&-&\frac{Ze}{m_d}E_x-\omega_dw_y-i\frac{k_{\perp}}{m_dn_{d0}}p_{\perp}^{\prime}-\nonumber\\
&-&\frac{ik_{||}}{m_d}(T_{||}-T_{\perp})\frac{B_x}{B_0},
\end{eqnarray}
\begin{equation}
\frac{\partial w_y}{\partial
t}=-\frac{Ze}{m_d}E_y+\omega_dw_x-\frac{ik_{||}}{m_d}(T_{||}-T_{\perp})\frac{B_x}{B_0},
\end{equation}
\begin{eqnarray}
\frac{\partial w_z}{\partial
t}=&-&\frac{Ze}{m_d}E_z-\frac{ik_{\perp}}{m_d}(T_{||}-T_{\perp})\frac{B_x}{B_0}-\nonumber\\
&-&\frac{ik_{||}}{m_dn_{d0}}p^{\prime}_{||},
\end{eqnarray}
where $\omega_d=ZeB_0/m_dc$ is the dust cyclotron frequency.

In order to determine the perturbations of pressure $p^{\prime}_{||}$ and 
$p^{\prime}_{\perp}$ we use the following equations 

\begin{equation}
\frac{\partial p^{\prime}_{||}}{\partial t}+p_{||}\nabla\cdot
\mathbf{w}+2p_{||}\nabla_{||}w_z=0,
\end{equation}
\begin{equation}
\frac{\partial p^{\prime}_{\perp}}{\partial t}+2p_{\perp}
\nabla\cdot \mathbf{w}-p_{\perp}\nabla_{||}w_z=0,
\end{equation}
equivalent to the Chew--Goldberger--Law (CGL) invariants, as well as the dust 
continuity equation 

\begin{equation}
\frac{\partial}{\partial
t}\left(\frac{\widetilde{n_d}}{n_{d0}}\right)+\frac{\partial
w_x}{\partial x}+\frac{\partial w_z}{\partial z}=0.
\end{equation}
It has to be mentioned that the pressure components $p_{||}$, $p_{\perp}$ 
in equations (22) and (23) are connected with both variation of density and 
perturbation of temperature -- the latter having adiabatic character in CGL 
theory. As a result, the equation for $\rm{curl}\,\vv{B}$ in linear approximation 
is written as 

\begin{equation}
(\nabla\times \mathbf{B})_{x,y}=\frac{4\pi e}{c}(n_{i0}V_{x,y}-n_{e0}v_{x,y}-Zn_{d0}w_{x,y}).
\end{equation}

\section{THE DISPERTION EQUATION}

\subsection{Derivation of the dispersion equation}

The equations~(10)--(12), (14)--(17), (19)--(25) and (7) represent a full set 
of equations of our problem. In general, this system of equations described the 
two types of low-frequency electromagnetic waves: Alfv\'en and magnetosound. 
Our goal is to demonstrate that behind the shock wave mirror instability 
develops in the dust component. It is well known that mirror instability is 
connected with magnetosound electromagnetic field with the wavevector complanar 
to the external magnetic and the wave magnetic fields. According to our 
choice $k_{\perp}$ points along~$x$-axis. Hence only the field components 
$E_y$, $B_x$ (and $B_z$) are connected with the magnetosound wave, while 
the components $E_x$ are $B_y$ small. In such conditions the 
solution of the motion equations~(10)--(11), (14)--(15) and (19)--(20)  for 
the Fourier components $\sim\exp i(k_{\perp} x+k_{||}z-\omega t)$ in the 
drift approximation ($\omega\ll\omega_i, \omega_d$) can be written as 

\begin{equation}
v_x=c\frac{E_y}{B_0},\,
v_y={-}\frac{c}{B_0}\left({E_x}-\frac{k_{\perp}}{k_{||}}{E_z}\right),
\end{equation}
\begin{equation}
V_x=c\frac{E_y}{B_0},\,
V_y={-}\frac{c}{B_0}\left(E_x+i\frac{\omega}{\omega_i}E_y\right),
\end{equation}
\begin{eqnarray}
w_x&=&c\frac{E_y}{B_0}, \nonumber\\  w_y={-}\frac{c}{B_0}\left(E_x-i\frac{\omega}{\omega_d}E_y\right)&-&\frac{ik_{||}c}{ZeB_0^2}(T_{||}-T_{\perp})B_x-\frac{ik_{\perp}
c}{ZeB_0n_{d0}}p^{\prime}_{\perp},  
\end{eqnarray}
\begin{eqnarray}
V_z={-}\frac{ie}{M\omega}E_z,\,\,w_z=\frac{k_{\perp}}{m_d\omega}(T_{||}-T_{\perp})\frac{B_x}{B_0}+\frac{k_{||}}{m_d\omega
n_{d0}}p^{\prime}_{||},
\end{eqnarray}
here the expression for $v_y$ is obtained with accounting~(13), while in the 
second equation of~(28) the term contained the field component $E_z$ is omitted 
as its contribution into the dispersion equation is small over $ZM/m_d$.

The equation for $\rm{curl}$ $\vv{E}$ gives the connection between the field 
components of interest

\begin{equation}
E_y=\frac{\omega}{k_{\perp} c}B_z,
  B_x={-}\frac{k_{||}}{k_{\perp}}B_z.
\end{equation}

From equations~(17) and (24) we have 
\begin{equation}
\frac{\widetilde{n_i}}{n_{i0}}=\frac{k_{\perp}
V_x+k_{||}V_z}{\omega}
\end{equation}
and 

\begin{equation}
\frac{\widetilde{n_d}}{n_{d0}}=\frac{k_{\perp}
w_x+k_{||}w_z}{\omega}.
\end{equation}
With using~(27), (29) and (30), it is convenient to rewrite equation (31) 
in the form 

\begin{equation}
\frac{\widetilde{n_i}}{n_{i0}}=\frac{B_z}{B_0}+\frac{ik_{||}e}{M\omega^2}E_z.
\end{equation}
In order to derive the perturbation of dust density let us write the equations 
following from~(22), (23) for the Fourier components: 

\begin{equation}
p_{||}^{\prime}=n_{d0}T_{||}\left(3\frac{\widetilde{n_d}}{n_{d0}}-2\frac{B_z}{B_0}\right),\,
p_{\perp}^{\prime}=n_{d0}T_{\perp}\left(\frac{\widetilde{n_d}}{n_{d0}}+\frac{B_z}{B_0}\right), 
\end{equation}
here equations (24), (29) and (30) are accounted. Then with accounting 
(28), (29) and (34) one obtains from (32) 

\begin{equation}
\frac{\widetilde{n_d}}{n_{d0}}=\frac{B_z}{B_0}
\left(1-\frac{k_{||}^2}{\omega^2}\frac{3T_{||}-T_{\perp}}{m_d}\right)\bigg/
\left(1-\frac{k_{||}^2}{\omega^2}\frac{3T_{||}}{m_d}\right).
\end{equation}

It is obvious that $T_{\perp}/m_d$ behind the shock front has the order of the 
square of the shock velocity, while $T_{||}/m_d$ characterizes the square 
of the thermal velocity of the dust component before the shock and is therefore 
negligibly small. Therefore the latter equation can be written as 

\begin{equation}
\frac{\widetilde{n_d}}{n_{d0}}=\frac{B_z}{B_0}\left(1+\frac{k_{||}^2T_{\perp}}{m_d\omega^2}\right).
\end{equation}
The field $E_z$ can be determined with using the quaineutrality equation 
$\widetilde{n_i}=\widetilde{n_e}+Z\widetilde{n_d}$ from (13), (33) and
(36):

\begin{equation}
E_z={-}i\frac{n_{e0}}{n_{i0}}\frac{M\omega^2}{e}\frac{k_{||}c_s^2}{\omega^2-k_{||}^2c_s^2}\frac{B_z}{B_0},
\end{equation}
where $c_s=(n_{i0}T_e/n_{e0}M)^{1/2}$ is the dust ion-acoustic speed. 
With using the latter equation let us write $v_y$ as 

\begin{equation}
v_y={-}c\frac{E_x}{B_0}-i\frac{n_{e0}}{n_{i0}}\frac{\omega^2}{\omega^2-k_{||}^2c_s^2}\frac{k_{\perp}
c_s^2}{\omega_i}\frac{B_z}{B_0}.
\end{equation}
The velocity $w_y$ can be found from~(28), (34) and (36) with 
accounting the above mentioned interrelation between the longitudinal and 
transverse dust temperatures:

\begin{eqnarray}
w_y=&-&\frac{c}{B_0}\left(E_x-i\frac{\omega}{\omega_d}E_y\right)-\nonumber\\
&-&i\frac{k_{\perp}
cT_{\perp}}{ZeB_0^2}\left(3+\frac{k_{||}^2T_{\perp}}{m_d\omega^2}\right)B_z.
\end{eqnarray}

Substituting (27), (38) and (39) into the $y$-projection of the 
$\rm{curl}\,\vv{B}$ we arrive at the dispersion equation 

\begin{eqnarray}
k^2V_A^2&=&\omega^2\left(1-\frac{\rho_i}{\rho}\frac{k_{\perp}^2c_s^2}{\omega^2-k_{||}^2c_s^2}\right)-\nonumber\\
&-&\frac{\rho_d}{\rho}\left(3+\frac{k_{||}^2V^2}{\omega^2}\right)k_{\perp}^2V^2.
\end{eqnarray}
Here $V_A^2=B_0^2/4\pi\rho$, $V^2=T_{\perp}/m_d$,
$\rho=\rho_i+\rho_d$, $\rho_i=Mn_{i0}$,~~$\rho_d=m_dn_{d0}$.

\subsection{The mirror instability}

In the absence of dust ($\rho_d=0$) equation (40) is biquadratic 

\begin{equation}
\omega^4-k^2(V_A^2+c_s^2)\omega^2+k^2k_{||}^2V_A^2c_s^2=0 ,
\end{equation}
and describes fast and slow magnetosonic waves. When dust is present equation 
~(40) is in general bicubic. Similar equation accounting kinetic effects is 
obtained and analyzed in details in~[16]. We consider here the most simple 
and obvious limiting case when the wave propagates almost perpendicular to the 
magnetic field: $k_{||}\ll k\approx k_{\perp}$ which corresponds to a 
long-wavelength modulation of the magnetosonic waves along the external 
magnetic field. Then with accounting the condition 
$\omega/k_{||}\gg c_s$ equation~(40) simplifies and becomes 

\begin{equation}
\omega^4-A\omega^2+B=0,
\end{equation}
where 

\begin{equation*}
A=k^2\left(V_A^2+\frac{\rho_i}{\rho}c_s^2+3\frac{\rho_d}{\rho}V^2\right),\nonumber\\
B=k^2k_{||}^2\left(V_A^2c_s^2+3\frac{\rho_d}{\rho}V^2c_s^2-\frac{\rho_d}{\rho}V^4\right).\nonumber
\end{equation*}
The solution of~(42) is 

\begin{equation}
\omega^2=\frac{1}{2}\left(A\pm\sqrt{A^2-4B}\right).
\end{equation}
The sign ``minus'' corresponds to the slow magnetosonic wave.
Unstable aperiodic solutions for such waves exist 
if $B<0$ which corresponds to the nonequality 

\begin{equation}
\frac{V^2}{c_s^2}>\frac{3}{2}+\sqrt{\frac{9}{4}+\frac{\rho}{\rho_d}\frac{V_A^2}{c_s^2}}.
\end{equation}
Immediately behind the shock where radiative losses are unimportant the 
parameter $V^2$, which characterizes the post-shock transversal dust 
temperature, is connected with the shock velocity by the relation 
$V^2=9v_s^2/32$, where $v_s$ is the shock velocity; here we assumed 
explicitle that a dust particle crossing the shock moves with respect to 
plasma with the velocity $3v_s/4$. The electron temperature immediately behind 
the front is $T_e=3m_ev_s^2/16$, i.e. $V^2/c_s^2=3m_i/2m_e$. Hence as might be 
thought the condition~(44) fulfils always as $m_i/m_e>\sqrt{\rho/\rho_d}$. 
In reality, however, $V^2/c_s^2=3/2$ is valid up to the region where radiative 
cooling sets on, since betatron acceleration responsible for an efficient dust 
destruction establishes far behind the front, where the electron-ion plasma 
is already isothermal. Therefore, the condition~(44) apparently does not 
fulfil.

In fact, it has to be taken into account that in the domain of radiative 
cooling all parameters included into~(44) vary. Namely, due to radiative 
cooling and an increase of magnetic field, and caused by it betatron 
acceleration the orbital velocity of dust particles $V$ grows as  
$V\propto \sqrt{B}$, or for a frozen-in magnetic field as 
$V\propto \sqrt{\rho}$. This gives for a cooling plasma where $\rho
T\simeq$ const $V^2/c_s^2\propto T^{-2}$; similarly in the radiation domain 
we have $V_A^2/c_s^2\propto T^{-2}$. Therefore, in the domain of radiative 
cooling the inequality~(44) can be written in the form 

\begin{equation}
\frac{3}{2}\frac{T_0}{T}>\sqrt{\frac{20}{M_A^2}\frac{\rho}{\rho_d}},
\end{equation}
where $T_0$ is the plasma temperature immediately behind the shock. 
Here immediately the shock front we explicitly used the relation 
$V_A^2/c_s^2\simeq 20/M_A^2$, where $M_A=v_s/V_{A0}$ is the Alfv\'en 
Mach number before the shock. Substituting here $T_0/T>10$ and a typical 
for the ISM value $\rho/\rho_d\simeq  100$, one can find the condition for 
the mirror instability $M_A>3$, which is deliberately valid for the shocks able 
to provide dust destruction.

\section{CONCLUSION}

In this paper we have considered a possibility for the mirror instability 
to develop in the dust component behind the shock. Flows behind the fronts 
of sufficiently strong shock waves ($M_A>3$) turn out to be unstable 
against the mirror instability. Consequently, a laminar magnetohydrodynamic 
flow behind the shock front supporting betatron acceleration of dust particles 
becomes violated. One can expect that it causes weakening of the betatron 
acceleration. Indeed, the mirror instability on nonlinear stages is known to 
result, in particular, in formation of magnetic voids, holes and growing 
loops swelling out behind the front to a few dust gyration radii~[17], which 
in our conditions is $r_{cd}\sim 0.3$~pc for typical parameters of dust 
particles and the shock velocity 100~km~s$^{-1}$. As a result, charged dust 
particles are expected to be expelled into the domains of weak magnetic field 
and decelerated as~$p^2_{\perp}\propto B$.

It follows from~(43) that the characteristic time of the instability has an 
order 

\begin{equation} 
\tau_m\sim \sqrt{\frac{\rho}{\rho_d}\frac{\lambda}{v_s}}, 
\end{equation}
where $\lambda$ is the perturbation wavelength. Assuming the wavelength to be 
equal to the thickness of the adiabatic SN remnant $\lambda\sim\Delta R\sim 0.1R_s$ 
and $\rho/\rho_d$ equal to a typical in the ISM value~100, one can find 
$\tau_m$ to be of the order of the remnant age. This estimate is an upper 
limit of the growth time -- short-wavelength perturbations grow obviously 
faster. The perturbation wavelength is restricted from below by the 
kinetic limit and by the size of the nonuniformity of the post-shock flow. 
The latter corresponds to the characteristic cooling length. The radiative 
cooling time can be estimated as $\tau_c\sim kT/\Lambda n \sim 10^{11}$~s 
for temperature $T\sim 3\times 10^5$ K, typical for the beginning of 
the radiative stage, and density $n\sim 1$~cm$^{-3}$; the corresponding 
scale is $\lambda\sim 0.3$~pc. The size of the remnant with the expansion 
velocity $\sim$100~km~s$^{-1}$ in a medium with mean density $n\sim
1$~cm$^{-3}$ is $30$~pc, hence the lower estimate of the characteristic 
time of the mirror instability on these stages is of 1$\%$ the remnant age. 
It can be expected therefore that the instability being developed in the whole 
possible wavelength range will efficiently restrict the rate of dust destruction. 
As mentioned above that at nonlinear stages the mirror instability results 
in breaking up of a regular flow with magnetic field parallel to the front, and 
leads essentially to formation of loop-like domains with a decreased magnetic 
field -- magnetic voids~[17]. It allows to estimate the factor of weakening of 
dust destruction assuming that magnetic field behind the shock does not increase 
due to radiative cooling, but in average remains rather the same as if 
the compression due to cooling would be absent. The corresponding 
decrease of the destruction efficiency would become an order of magnitude 
in the range of the ``danger'' velocity interval 100 km~s$^{-1}<v_s<300$~km~s$^{-1}$. 
Firm estimates can be obtained only from numerical modeling of the instability. 

This work is supported by the Russian Foundation of Basic Reserach 
(project codes 08-02-00933 and 09-02-97019), and in part by Deutsche Forschungs 
Gemeinschaft under the project SFB 591.

\end{document}